\documentclass[conference]{IEEEtran} 
\IEEEoverridecommandlockouts

\usepackage[utf8]{inputenc}
\usepackage[T1]{fontenc}
\usepackage{siunitx}
\usepackage{graphicx}
\usepackage{amsmath, amssymb, bm}
\usepackage{booktabs}
\usepackage{url}
\usepackage{balance}
\usepackage{hyperref}

\title{The SpinQuest Microwave System for Dynamic Nuclear Polarization}

\author{%
\IEEEauthorblockN{Vibodha Bandara\IEEEauthorrefmark{1}\IEEEauthorrefmark{2},
Jordan D. Roberts\IEEEauthorrefmark{1},
Dustin Keller\IEEEauthorrefmark{1}}
\IEEEauthorblockA{\IEEEauthorrefmark{1}University of Virginia, Charlottesville, VA, USA}
\IEEEauthorblockA{\IEEEauthorrefmark{2}Department of Instrumentation and Automation Technology, University of Colombo, Sri Lanka}
\thanks{This work was supported by the U.S. Department of Energy, Office of Nuclear Physics (insert award numbers).}
}

\begin{document}
\maketitle

\begin{abstract}
The SpinQuest experiment at Fermilab uses a dynamically polarized solid-ammonia target to probe the spin structure of the proton. Stable, optimized microwave-driven dynamic nuclear polarization (DNP) is therefore required under cryogenic, high-field, and high-radiation conditions. We present the design, operation, and automation of a $\sim$140~GHz microwave system based on an extended interaction oscillator (EIO). The source is integrated with motorized cavity tuning, real-time polarization feedback from a continuous-wave NMR system, cryogenic diagnostics, and remote operation to reduce radiation exposure to sensitive control electronics.

To enable continuous optimization of target polarization, we developed an automation framework supported by a Monte Carlo target-control emulator, or digital twin, of the DNP process. The simulation incorporates rate-equation dynamics, frequency-dependent steady-state behavior, dose-induced frequency drift, heat-induced depolarization, and realistic NMR measurement noise. This framework was used to design and benchmark control strategies, including a heuristic feedback algorithm, reinforcement learning (RL), and unsupervised RL approaches. These methods enable autonomous frequency tuning, improve polarization ramp-up efficiency, and maintain near-optimal polarization as target conditions evolve.

We also demonstrate the integration of EIO power-supply control into the feedback framework through modulation of the anode voltage. This provides an additional actuator for simultaneous control of microwave frequency and RF power. Combined control of cavity position and anode voltage allows the system to avoid frequency-dependent power nonuniformities and to better address broad Larmor-frequency distributions in irradiated targets.

The results establish a scalable framework for data-driven control of complex microwave systems in polarized-target experiments, with potential applications in future high-precision spin-physics measurements and other cryogenic, high-field systems.
\end{abstract}

\section{Introduction}
\label{sec:introduction}

The SpinQuest experiment at Fermi National Accelerator Laboratory (Fermilab) is designed to investigate the spin structure of the proton (and neutron) \cite{SpinQuest}. The experiment uses a 120~GeV proton beam from the Main Injector incident on a solid polarized-ammonia target, which is a central component of the apparatus.

Target polarization is achieved through dynamic nuclear polarization (DNP), which has been extensively described by Abragam \cite{Abragam, abragam2, abragam_cea, ABRAGAM1962310}, Borghini \cite{MB, Borghini2}, and Meyer and Crabb \cite{crabb1997solid}. In DNP, microwave irradiation drives coupled transitions involving nuclear spins and unpaired electron spins. Under suitable conditions, the much larger electron polarization is transferred to the nuclear spin system, greatly enhancing the nuclear polarization \cite{Tapio}.

Following Overhauser's theoretical prediction of enhanced nuclear polarization in non-metals \cite{Overhaus} and Abragam's experimental confirmation \cite{Abragam}, DNP became the standard method for producing polarized solid targets. Early implementations used relatively low-power microwave sources, including klystrons at the National Laboratory for High Energy Physics (KEK) operating at approximately 500~mW \cite{KEK} and carcinotrons at Brookhaven National Laboratory operating near 3~W \cite{bnl}.

Subsequent advances in millimeter-wave technology enabled higher output power and improved frequency control through the use of extended interaction oscillators (EIOs) \cite{Day1966}. These sources increased achievable polarization and reduced polarization ramp-up time. The foundational work by D. G. Crabb demonstrated high polarization in solid-state ammonia targets \cite{PhysRevLett.64.2627}, followed by polarized-target experiments such as E143 \cite{E143} and E155 \cite{E155} at the Stanford Linear Accelerator Center (SLAC), and several Jefferson Lab experiments \cite{clas,g2p,APOLO,E03,e06,E93,SANE,EG1,RAS}.

The EIO provides higher electronic efficiency and output power than alternatives such as klystrons \cite{Day1966}. It can be configured over a range of frequency and power settings through the power supply, while fine frequency control is achieved by mechanically changing the cavity dimensions with a motorized tuning shaft. The system response is monitored with an output frequency counter and a continuous-wave nuclear magnetic resonance (NMR) system that measures the target polarization.

For highly ionizing beams at intensities above $3\times10^{12}$ protons/s, as in SpinQuest, radiation-tolerant control systems with remote operation and automated microwave feedback are essential. Semiconductor instrumentation must be routed approximately 20~m outside the target enclosure. Radiation levels near the target can reach total ionizing doses on the order of $10^{3}$~Gy(Si), which is sufficient to cause severe damage to sensitive electronics \cite{semirad}. In addition, continuous microwave-frequency tuning is required to maintain optimal polarization during irradiation. This requirement arises primarily from radiation-induced evolution of paramagnetic centers in the target material, which modifies the electron spin resonance conditions and shifts the optimal DNP frequency as a function of accumulated dose.

To address these challenges, SpinQuest uses a $\sim$140~GHz EIO to drive DNP in an ammonia target operated near 1~K in a homogeneous 5~T magnetic field. The EIO is equipped with a stepper-motor-controlled cavity tuner and an electronically regulated power supply, enabling precise control of both frequency and output power. The control electronics and associated instrumentation are located remotely, outside the high-radiation environment, while the EIO source is positioned close to the cryostat and target insert to minimize microwave transmission losses. The DNP microwave-frequency control system can be operated either manually or automatically. To the best of our knowledge, this work represents the first automated control of a DNP microwave system in which the paramagnetic complex evolves during irradiation.

The remainder of this paper is organized as follows. Section~\ref{sec:eio} describes the EIO source and power conditioner. Section~\ref{sec:dnp} summarizes the DNP process relevant to the control problem. Section~\ref{sec:microwave} describes the SpinQuest microwave transmission and frequency-control system. Section~\ref{sec:automation} presents the automation strategy and simulation framework. Section~\ref{sec:conclusion} summarizes the main results and conclusions.

\section{Extended Interaction Oscillator}
\label{sec:eio}

Microwave radiation was generated with a continuous-wave EIO manufactured by Communications \& Power Industries Canada Inc. The device used in this system was a CPI model VKT2438P6M oscillator, serial number E1289F5. The VKT2438P series is a motor-tuned millimeter-wave oscillator family intended for F-band radar and scientific instrumentation. CPI specifies the VKT2438P model series as having a nominal center frequency of \SI{140}{\giga\hertz}, continuous-wave output power on the order of \SI{20}{\watt}, a mechanical tuning range of approximately \SI{\pm 1.0}{\giga\hertz}, and a typical electronic tuning range of around \SI{400}{\mega\hertz} \cite{ORC1968}.

An EIO is a linear-beam vacuum-electron microwave source that converts electron-beam kinetic energy into coherent RF power. Electrons are emitted from a heated thermionic cathode and accelerated by the applied cathode-to-anode voltage. The beam is focused through an interaction region containing a resonant slow-wave structure. In extended-interaction devices, the RF circuit contains multiple interaction gaps within each cavity, commonly implemented with a ladder-type geometry. This extended interaction increases the effective coupling between the electron beam and the RF field, improving beam modulation and energy transfer at millimeter-wave frequencies \cite{steer_eik_technology}.

During operation, the resonant electromagnetic field velocity-modulates the electron beam, causing electron bunching as the beam propagates through the interaction region. When the bunched beam passes through the appropriate phase of the RF field, part of its kinetic energy is transferred to the cavity mode, sustaining oscillation and producing continuous-wave millimeter-wave output. The generated microwave power is coupled out of the tube through a waveguide, and the spent electron beam is collected in the collector. Body current is monitored because excessive body current indicates unwanted beam interception by the tube structure and can be used as an indicator of tube alignment, loading, or operating condition.

Factory test data for the VKT2438P2 tube used here show operation over \SIrange{138.5}{141.5}{\giga\hertz}. Over this range, the measured RF output power was approximately \SIrange{16}{19}{\watt}. The listed operating conditions included a cathode voltage of approximately \SI{9.765}{\kilo\volt}, cathode current of \SI{85}{\milli\ampere}, anode voltage near \SI{6.67}{\kilo\volt}, heater voltage of \SI{6.3}{\volt}, and deionized-water cooling at \SI{1.5}{\liter\per\minute} \cite{cpi_test_report_2016}. These values are consistent with the high-voltage, liquid-cooled requirements of compact continuous-wave millimeter-wave vacuum-electron sources.

The EIO is operated using a CPI VPW2827 electronic power conditioner. This unit provides the regulated electrical supplies and protection circuitry required for safe operation of CPI continuous-wave extended-interaction klystrons and oscillators. The power conditioner supplies the cathode, anode, collector, and heater operating conditions and monitors critical tube currents. CPI specifies the VPW2827 for beam voltages between \SI{-4.0}{\kilo\volt} and \SI{-12.0}{\kilo\volt}, anode voltages from \SI{1.0}{\kilo\volt} to \SI{6.0}{\kilo\volt}, heater voltages between \SI{-5.0}{\volt} and \SI{-7.0}{\volt}, collector currents up to \SI{130}{\milli\ampere}, and body-current monitoring up to \SI{20}{\milli\ampere} \cite{EIO_Manual}.

The power conditioner also performs the required start-up sequencing and fault-protection functions. In a typical operating sequence, the heater supply is enabled first so that the cathode reaches the required emission temperature. After the cathode is conditioned, the high-voltage supplies are enabled under interlock control, establishing the beam, anode, and collector potentials needed for oscillation. The VPW2827 includes an external interlock input that inhibits high voltage and provides fault protection for designated CPI EIK and EIO tubes. CPI also notes that operating parameters are programmed into on-board memory and password protected, allowing the power conditioner to be matched to a specific tube \cite{EIO_Manual}.

Together, the CPI VKT2438P2 EIO and VPW2827 electronic power conditioner \cite{PS_Manual} form a compact continuous-wave millimeter-wave source. The EIO provides the resonant vacuum-electron interaction required to generate microwave power near \SI{140}{\giga\hertz}, while the power conditioner provides the regulated high-voltage, heater, monitoring, and protection functions required for stable and repeatable tube operation.

\section{Dynamic Nuclear Polarization}
\label{sec:dnp}

DNP enhances nuclear spin polarization through an applied microwave field that couples the electron and nuclear spin systems. Several mechanisms can contribute to polarization transfer, depending on the material and experimental conditions. The primary mechanisms include the Overhauser effect (OE), solid effect (SE), cross effect (CE), and thermal mixing (TM) \cite{Tapio}. In this work, the solid effect is adopted as a simplified framework \cite{Tapio}; in practice, however, the dominant mechanism is determined by the electron spin resonance (ESR) linewidth, radical species, microwave power, and microwave frequency.

In the solid effect, paramagnetic radicals embedded in the target material provide unpaired electron spins that become highly polarized at high magnetic field and low temperature. Microwave irradiation at frequencies near the sum or difference of the electron and nuclear Larmor frequencies drives coupled electron--nuclear transitions through the interaction of the spin operators $\mathbf{S}$ and $\mathbf{I}$ \cite{SE}. These interactions give rise to nominally forbidden transitions, most notably simultaneous electron and nuclear spin flips (flip--flop transitions), while flip--flip transitions are strongly suppressed by the corresponding selection rules \cite{jeff1}.

Efficient DNP requires two conditions: (i) a large disparity between the electron and nuclear gyromagnetic ratios, $\gamma_e \gg \gamma_n$, and (ii) a relaxation-time hierarchy in which the electron spin--lattice relaxation time is much shorter than that of the nuclei, $T_{1e} \ll T_{1n}$. Together, these conditions allow the electron spin system to reach high polarization and rapidly re-equilibrate, thereby acting as a continuously replenished polarization reservoir for the nuclear spins.

Under thermal equilibrium (TE), spin polarization originates from the Boltzmann population difference between Zeeman-split energy levels. For a spin-$\tfrac{1}{2}$ system in a magnetic field $B_0$, the polarization is
\begin{equation}
P = \tanh\left(\frac{\gamma \hbar B_0}{2 k_B T}\right) \approx \frac{\gamma \hbar B_0}{2 k_B T},
\label{Polarization}
\end{equation}
where the approximation holds in the high-temperature limit $\gamma \hbar B_0 \ll k_B T$. The linear dependence on the gyromagnetic ratio $\gamma$ implies that, under identical field and temperature conditions, electron spins attain polarizations that are orders of magnitude larger than those of nuclear spins. This intrinsic polarization imbalance, combined with favorable relaxation dynamics, enables efficient transfer of polarization from the electron spin system to the nuclear spin system.

When the hierarchy $T_{1n} \gg T_{1e}$ is satisfied, electron spins repolarize rapidly while nuclear spins retain the transferred polarization for much longer times. Substantial nuclear polarization can therefore accumulate under continuous microwave irradiation. A useful measure of the efficiency of this process is
\begin{equation}
f = \frac{N_n T_{1e}}{N_e T_{1n}} \ll 1,
\label{f}
\end{equation}
where $N_n$ and $N_e$ are the number densities of nuclear and electron spins, respectively. This condition expresses the requirement that the electron spin reservoir be both sufficiently abundant and sufficiently fast-relaxing to sustain polarization transfer to the more slowly relaxing nuclear system.

The relative abundance of nuclei to paramagnetic centers is characterized by
\begin{equation}
C = \frac{N_n}{N_e},
\label{C}
\end{equation}
which links the spin density of the host material to that of the radical population responsible for DNP\@. The achievable polarization is therefore controlled by the relaxation hierarchy and by the balance between nuclear spin density and effective electron-spin-center concentration. In practice, polarization is generated locally near the paramagnetic centers and then propagated through the material by nuclear spin diffusion.

This balance evolves continuously under beam exposure. Radiation damage modifies the microscopic defect structure of the material, alters the radical population, and can create additional depolarizing centers. Thus, irradiated material contains effective polarization sources associated with active paramagnetic centers that support DNP and polarization sinks associated with beam-induced damage that reduces transfer efficiency or enhances local relaxation. The net polarization profile is determined by the competition among microwave-driven polarization transfer, spin diffusion, and radiation-induced depolarization. As the accumulated dose increases, these effects can shift the optimal DNP conditions and reduce the overall effectiveness of the polarization process.

\section{The SpinQuest Microwave System}
\label{sec:microwave}

\subsection{Microwave System Overview}

The EIO describe in Section \ref{sec:eio} generates microwaves near 140~GHz, although the accessible frequency range can be modified for best performance over the frequency domain of interest using the power supply settings. Frequency tuning during experimental operation is achieved through two complementary mechanisms. Coarse adjustment is obtained by varying the tube operating anode voltage, which changes the electron-beam energy and therefore the synchronism condition between the beam and the resonant electromagnetic mode. This shifts the frequency at which efficient power transfer occurs within the interaction structure. Fine tuning is provided by mechanical adjustment of the resonant cavity dimensions through rotation of the EIO tuning shaft. Changing the cavity geometry modifies the electromagnetic boundary conditions and hence the resonance frequency of the oscillator mode. Together, these electrical and mechanical controls provide the frequency agility required to track the evolving DNP optimum during target irradiation.

The EIO was configured to produce less than 20~W to meet Fermilab safety requirements \cite{ORC1968}. From the generator the microwaves are produced at D-band through a WR-6 rectangular waveguide. The output is first routed through a straight guide section into a 40~dB directional coupler. The coupled arm is used for frequency diagnostics: it is connected to a harmonic mixer, and the resulting intermediate-frequency signal is carried through a 25~m Gore cable to a frequency counter for accurate measurement of the source frequency. The main transmission path continues through the direct arm of the coupler into a D-band waveguide bend that redirects the microwave field downward. A WR-6-to-WR-15 taper then interfaces the D-band section to the V-band transport line. Downstream of this transition, a fixed 10~dB attenuator reduces the transmitted power before a tapered mode converter transforms the rectangular TE$_{10}$ mode into the circular TE$_{11}$ mode required for propagation in the final circular guide. The microwave power is then carried through a straight circular waveguide to the target insert.

A remotely adjustable attenuator configuration was also explored for active regulation of the microwave power delivered to the DNP target. This configuration used a Mi-Wave Model 520V/385 variable attenuator (50--75~GHz, nominal attenuation range 0--25~dB), allowing the delivered target power to be adjusted over approximately 0--2.5~W. This capability allows elevated power during initial polarization buildup, when faster DNP can be advantageous, followed by reduced power once the target approaches its operating polarization. The reduced-power mode mitigates microwave-induced helium boil-off and decreases the steady-state thermal load. In conventional DNP target systems usable microwave power is often limited to $\sim$1~W by the available cooling power. In contrast, the SpinQuest target system has substantially higher refrigeration capacity \cite{asmefridge}, allowing higher-power operation during polarization ramp-up while still benefiting from reduced-power running to conserve helium and minimize cryogenic load.

\begin{figure}[h!]
    \centering
    \includegraphics[width=9cm]{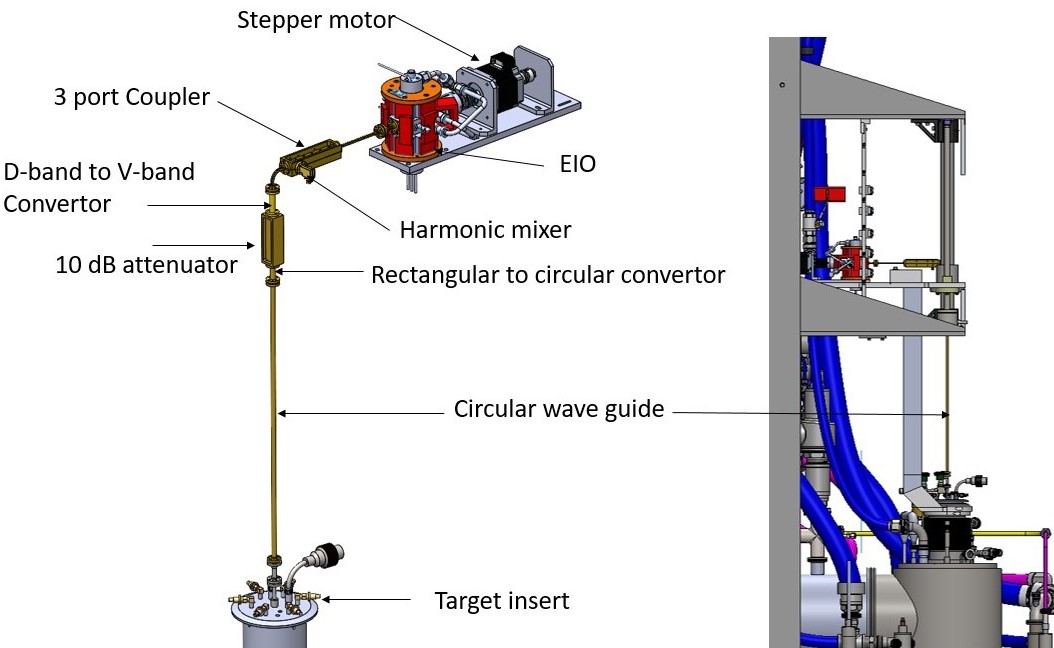}
    \caption{EIO microwave generator mounted on the moving table of the target lifter. Waveguides connect the EIO to the target insert and guide the microwave signal required for DNP\@.}
    \label{fig:wg_set}
\end{figure}

Figure~\ref{fig:wg_set} shows the waveguide arrangement in the microwave system and the placement of the assembly on the target lifter. The lifter selects which target cell is positioned in the proton beam.

Microwave power propagating through the waveguide chain is attenuated primarily by conductor loss, $\alpha_c$, and dielectric loss, $\alpha_d$:
\begin{equation}
\alpha_{\mathrm{tot}} = \alpha_c + \alpha_d.
\end{equation}
At 140~GHz, conductor loss is generally the dominant contribution in metallic waveguides and is governed by the surface resistance of the guide walls. This behavior is set by the electromagnetic skin depth, $\delta_s$, defined as the depth at which the field amplitude inside the conductor falls to $1/e$ of its surface value:
\begin{equation}
\delta_s = \sqrt{\frac{1}{\pi f \mu \sigma}},
\label{equ:skin}
\end{equation}
where $f$ is the microwave frequency, $\mu$ is the magnetic permeability, and $\sigma$ is the electrical conductivity of the wall material. The corresponding surface resistance is
\begin{equation}
R_s = \frac{1}{\sigma \delta_s}
    = \sqrt{\frac{\pi f \mu}{\sigma}}.
\label{equ:Rs}
\end{equation}

For the dominant TE$_{10}$ mode in a rectangular waveguide, the conductor attenuation may be written as
\begin{equation}
\alpha_c =
\frac{R_s}{a^3 b \beta k \eta}
\left(2 b \pi^2 + a^3 k^2 \right),
\label{equ:conloss}
\end{equation}
where $a$ and $b$ are the broad and narrow waveguide dimensions ($a>b$), $k=2\pi/\lambda$ is the free-space wavenumber, $\eta=\sqrt{\mu/\varepsilon}$ is the intrinsic impedance of the medium, and $\beta$ is the propagation constant. For TE$_{10}$ propagation,
\begin{equation}
\beta = \sqrt{k^2-\left(\frac{\pi}{a}\right)^2}.
\label{equ:rec_beta}
\end{equation}
Equations~\eqref{equ:skin}--\eqref{equ:conloss} show that increasing frequency decreases skin depth and increases surface resistance, leading to larger conductor losses. High-conductivity materials are therefore preferred in rectangular transmission-line sections. Copper is commonly used because of its high conductivity and ease of fabrication, while silver and gold plating can further reduce surface resistance. Silver provides the highest electrical conductivity, whereas gold is often favored because of its chemical stability and resistance to oxidation.

Dielectric loss is the second contribution to the total attenuation and arises from dissipation in the medium filling the waveguide. For a low-loss dielectric, this contribution can be written as
\begin{equation}
\alpha_d = \frac{k^2 \tan\delta}{2\beta},
\label{equ:dieloss}
\end{equation}
where $\tan\delta$ is the dielectric loss tangent. Larger values of $\tan\delta$ correspond to greater dissipation. In air-filled or evacuated waveguide sections, however, $\tan\delta$ is very small, and dielectric loss is typically negligible compared with conductor loss.

The downstream portion of the microwave transport line uses circular waveguide operating in the dominant TE$_{11}$ mode. In this application, circular waveguide is advantageous for the long transport section because it provides rotational symmetry, facilitates mechanical alignment and repeated assembly, and supports efficient high-frequency transmission with acceptable loss over meter-scale distances. The absence of sharp corners also reduces wall-current crowding relative to practical rectangular-guide implementations. These features make circular waveguide well suited for transporting high-frequency microwave power from the source region to the cryogenic target insert.

In the present system, the circular waveguide is fabricated from copper--nickel (CuNi) tubing. Although CuNi has lower electrical conductivity than pure copper or plated waveguide, it provides a favorable balance among microwave performance, mechanical robustness, and environmental stability. CuNi retains good structural integrity under cryogenic conditions and is less susceptible to mechanical degradation during thermal cycling, which is important for operation in a polarized-target cryostat. Although its conductivity does not improve as strongly upon cooling as that of high-purity copper, the resulting attenuation remains acceptable for the present geometry and operating frequency. Additional internal silver or gold plating was therefore not required. The calculated attenuation for the CuNi circular waveguide used in this system is approximately 1~dB/m at 140~GHz.

\begin{figure}[h!]
    \centering
    \includegraphics[width=5cm]{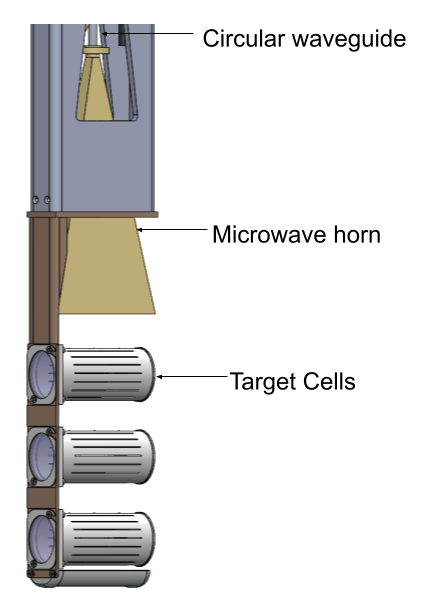}
    \caption{Lower portion of the target insert showing the circular-waveguide termination and microwave horn positioned above the target cells.}
    \label{fig:horn}
\end{figure}

\begin{figure}[h!]
    \centering
    \includegraphics[width=11cm]{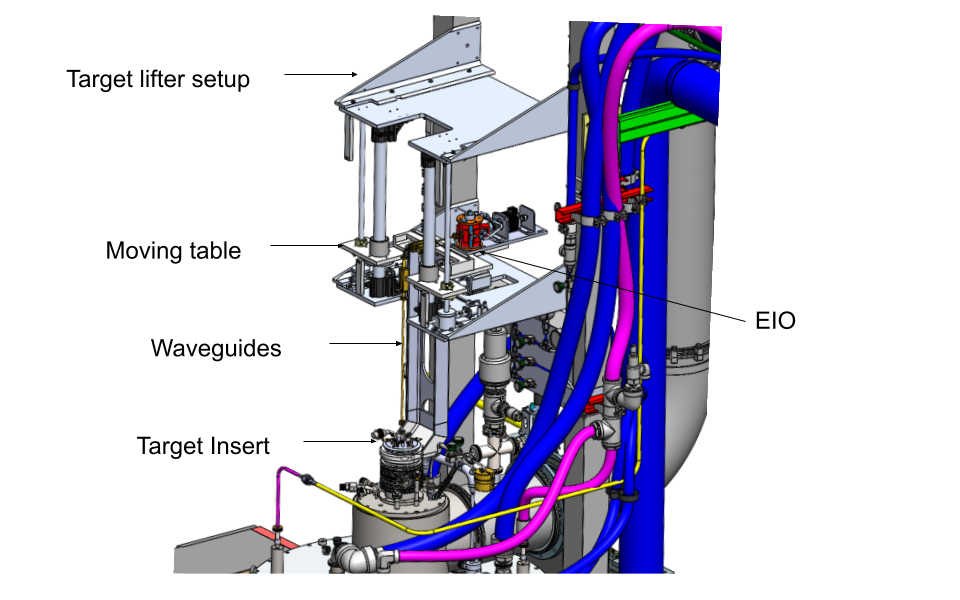}
    \caption{EIO mounted on the target lifter, with microwave output coupled to the target insert through waveguides.}
    \label{fig:EIO_lifter}
\end{figure}

The circular waveguide inside the target insert is 1.33~m long and terminates in a microwave horn positioned above the target cells, where the transmitted power is radiated onto the target material, as illustrated in Fig.~\ref{fig:horn}. The horn is geometrically configured to radiate the full profile of the 80 mm by 25 mm target material cell. The lower section of the insert includes a curved reflective surface designed to redirect microwaves that are not absorbed by the resonant target material back toward the target cells. Figure~\ref{fig:EIO_lifter} shows the microwave assembly and the top of the DNP cryostat.

\subsection{Frequency-Control Unit}

\begin{figure}[h!]
    \centering
    \includegraphics[width=9cm]{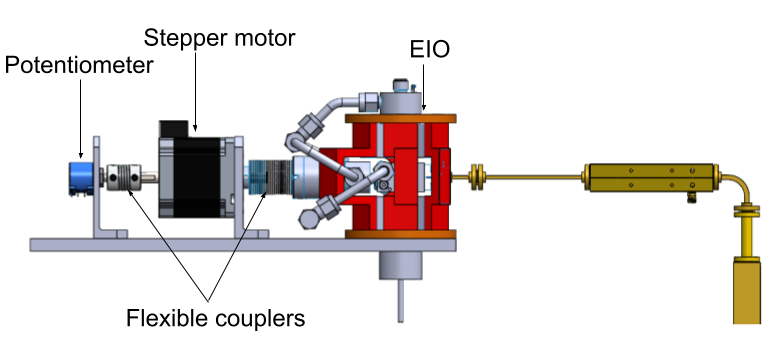}
    \caption{Motorized actuator assembly connected to the EIO for precise frequency tuning. The stepper motor, flexible coupler, and position-feedback potentiometer are mounted on the EIO assembly.}
    \label{fig:microwave_motor}
\end{figure}

The rotation of the shaft, which changes the resonant dimensions of the interaction structure and shifts the oscillation frequency, is done with a custom control configuration setup to meet the experiments needs. The actuator is controlled through a LabVIEW interface and is driven by a five-phase stepper motor for precise, repeatable positioning. The motor has 1000 steps per revolution, corresponding to an angular resolution of $0.36^\circ$ per full step. Mechanical coupling between the motor shaft and the EIO tuning shaft is provided by a flexible coupler, which compensates for small alignment errors and reduces stress on the drive assembly. Absolute position feedback is obtained from a potentiometer mounted on the rear shaft of the motor, providing reproducible position readback for automation and closed-loop operation. The EIO tuning shaft permits a total travel of four turns and is protected by hard stops that prevent overtravel. This range of cavity adjustment provides an overall microwave-frequency tuning span of approximately 3\%. Figure~\ref{fig:microwave_motor} shows the motorized tuning assembly attached to the EIO.

The EIO output frequency is measured with an EIP frequency counter connected to the harmonic mixer in the EIO waveguide assembly. In the nominal operating configuration, the counter is located remotely about 20 m away outside of the target alcove. This read-back was not always reliable because the long cable introduced additional attenuation in the intermediate-frequency signal path. To improve the signal quality during calibration, the frequency counter was temporarily installed inside the target cave and connected to the mixer with a 2~m Gore cable for calibration and frequency mapping.

\begin{figure}[!htbp]
    \centering
    \includegraphics[width=9cm]{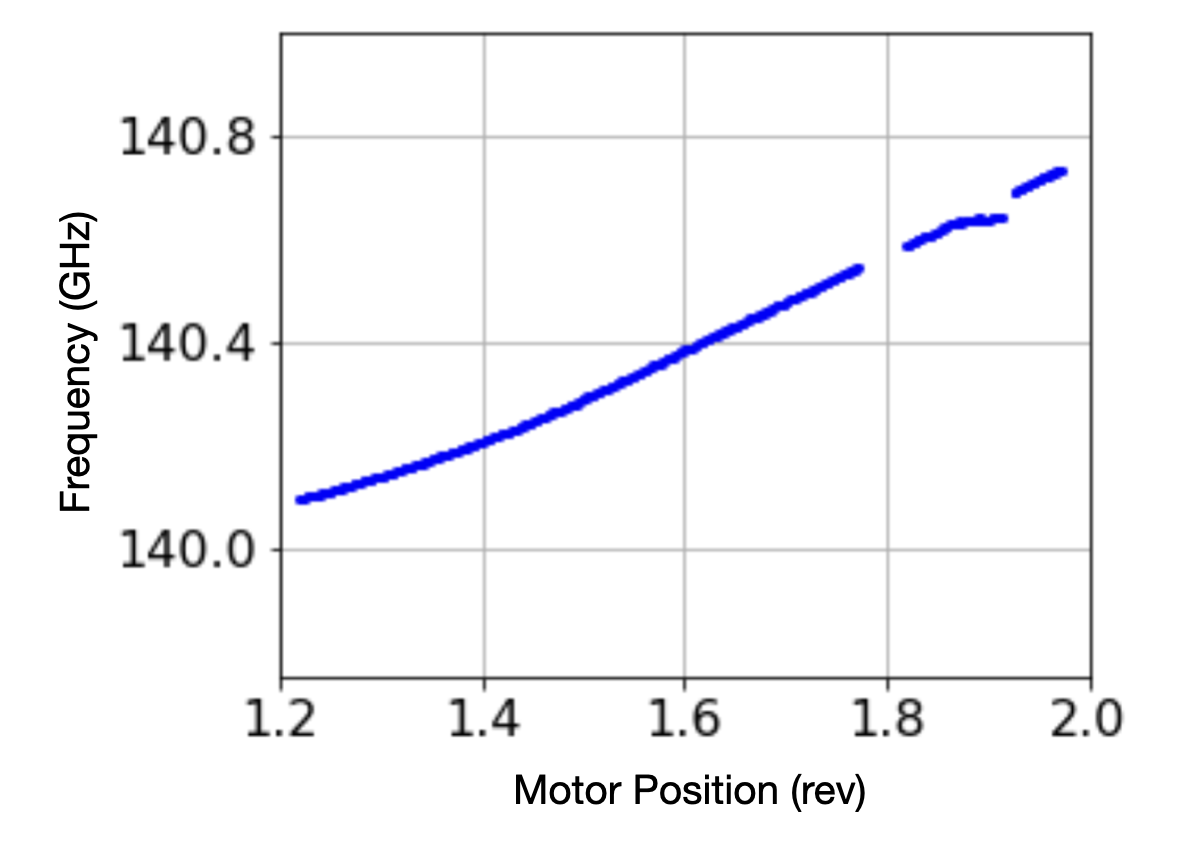}
    \caption{Measured EIO output frequency as a function of tuning-shaft position. Missing portions of the curve near 1.8 and 1.95 revolutions correspond to microwave power pockets where the frequency-counter readout was unreliable.}
    \label{fig:microwave_fre_plot}
\end{figure}

A calibration of output frequency versus tuning-shaft position was performed by stepping the motor in two-step increments from the counterclockwise (CCW) mechanical limit to the clockwise (CW) limit. This scan was repeated multiple times to evaluate reproducibility. The EIO can exhibit power pockets, where the microwave output is too low to deliver adequate power to the horn or too weak to be detected reliably by the EIP frequency readout. Figure~\ref{fig:microwave_fre_plot} shows an example of frequency versus motor position with this type of discontinuity. In this case, power pockets are observed near 1.8 and 1.95 revolutions. Small adjustments to the EIO power-supply settings were made until a reproducible frequency readout was obtained over the desired region. Approximately ten scans were then recorded. The measured frequency followed the same functional dependence on shaft position in each scan, demonstrating good repeatability of the mechanical tuning system. The resulting frequency--position relation was incorporated into the LabVIEW control software as a lookup table, allowing the system to determine the microwave frequency corresponding to a given motor position and to calculate the motor position required to reach a specified frequency.

The potentiometer connected to the motor assembly provides an output voltage that serves as an absolute measure of the tuning-shaft position. In the motor-control system, this potentiometer output is monitored by a window-comparator circuit built using operational amplifiers.

Target polarization is displayed as a function of time by polling the NMR control computer at 2~s intervals. When beam is delivered to the target, the LabVIEW virtual instrument (VI) also acquires beam information, integrates the incident beam intensity as a function of time, and displays the accumulated proton dose on the target in real time.

\begin{figure*}[!htbp]
    \centering
    \includegraphics[width=\textwidth]{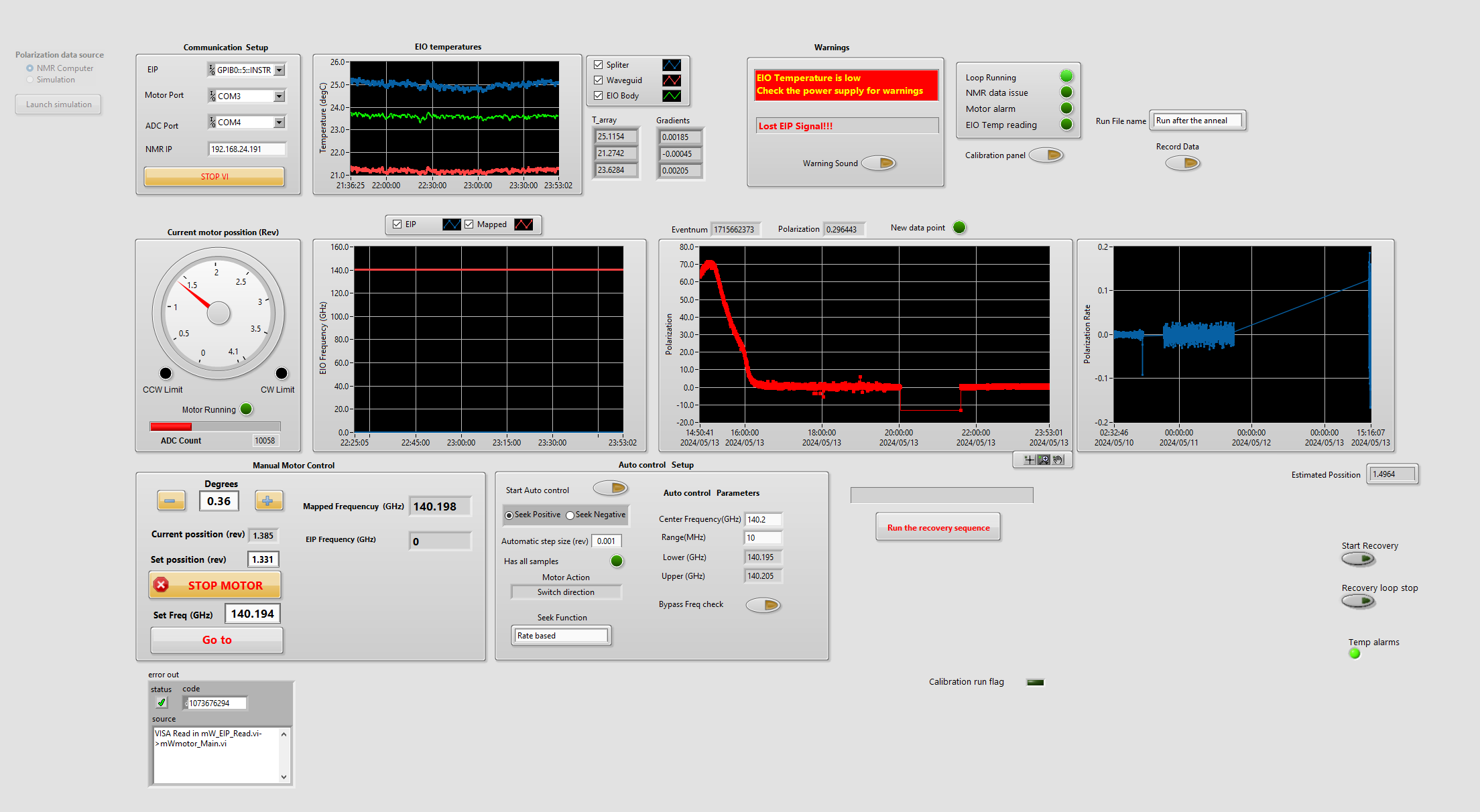}
    \caption{LabVIEW interface for microwave-frequency control and monitoring. The main strip charts display target polarization, microwave frequency, and accumulated beam intensity in real time. The interface allows both manual frequency adjustment and activation of the automatic optimization routine. Temperatures measured at selected locations on the EIO and waveguide assembly are also displayed.}
    \label{fig:microwave_VI}
\end{figure*}

The user may adjust the microwave frequency either by commanding the motor to a specified shaft position or by entering a desired frequency directly. In the latter mode, the LabVIEW VI (Fig.~\ref{fig:microwave_VI}) uses the frequency--position lookup table to determine the corresponding motor position and then drives the actuator to that setting. Additional fine tuning can be carried out through small incremental motor movements. In full-step operation, the size of the motor step ($0.36^\circ$ per step) corresponds to approximately 1~MHz change in output frequency near 140~GHz. The controller can also be operated in microstepping mode to increase the commanded positioning resolution, although the effective frequency resolution is ultimately limited by the mechanical response of the tuning assembly.

Three thermocouples were installed on the microwave system: one on the EIO body, one on the first output waveguide section, and one on the directional coupler. A flow sensor verifies continuous cooling-water flow through the microwave chiller; if flow is lost, the power supply automatically shuts off to prevent overheating of the EIO generator. In addition, the thermocouple mounted on the EIO body is incorporated into the hardware interlock and disables microwave power if the temperature exceeds $80\,^\circ\mathrm{C}$. The other two thermocouples provide operational diagnostics for the microwave source and transmission line.

During normal operation, the temperatures of the waveguide components typically rise above approximately $33\,^\circ\mathrm{C}$. An alarm condition is generated if these temperatures fall unexpectedly, which can indicate a power-supply trip, operation in a frequency power pocket, or a sudden fault that interrupts microwave output. The LabVIEW VI continuously monitors the thermocouple signals, and an unexpected drop in waveguide temperature is therefore used as an indirect indicator that microwave power has been interrupted, prompting operator intervention.

The automatic control mode implemented in the LabVIEW VI adjusts the motor position iteratively to maximize the measured target polarization, either positive or negative, depending on the selected operating condition. This optimization uses the frequency--position lookup table together with continuous polarization feedback from the NMR system to retune the EIO as target conditions evolve. In this way, the microwave frequency can be maintained near the polarization optimum throughout beam exposure, reducing the need for manual retuning and improving the stability of target operation. Frequency regulation can therefore be performed either manually by the operator or automatically through the control system.

\section{Automation}
\label{sec:automation}

The automated microwave control system must support two operating modes. First, it must identify the optimal microwave frequency for rapid buildup of either positive or negative polarization across different target materials and magnetic-field settings, typically 2.5 or 5~T. Second, it must continuously retune the microwave frequency during operation to compensate for irradiation-induced evolution of the paramagnetic complex and thereby maintain the maximum achievable polarization.

A control system that persistently tracks and optimizes the microwave frequency can improve the experimental figure of merit by maximizing target polarization throughout beam delivery and reducing the time required to recover optimal operating conditions.

To investigate candidate automation strategies, Monte Carlo simulations of the DNP process were developed. These simulations provide a flexible framework for studying the response of the polarized target system under varying conditions and serve as a platform for training and validating microwave-control algorithms.

\subsection{Simulations}

To develop and benchmark candidate automation algorithms in a controlled and repeatable framework, we constructed a Monte Carlo DNP target-control emulator, or digital twin, that captures the coupled response of the polarized target, NMR readout, and microwave-frequency control system. The framework can operate in several configurations: fully simulated NMR data with simulated microwave tuning, real NMR data with simulated microwave actuation, simulated NMR data driving the physical microwave tuning system, or fully integrated operation with both experimental NMR data and hardware control. In all cases, the automation software interprets the evolving NMR-derived polarization response and commands changes to the microwave tuning system, implemented as changes in the effective cavity size, to increase or decrease the operating frequency and track the optimal DNP condition as the target evolves. The emulator was designed to reproduce three key aspects of the experimental response: (i) the frequency-dependent steady-state polarization profile, or DNP S-curve; (ii) the finite ramp-up and decay dynamics following changes in microwave operating conditions; and (iii) the slow, dose-dependent evolution of the irradiated target, which causes both degradation of the achievable polarization and drift of the optimal microwave frequency.

An important design objective was to generate synthetic data that are indistinguishable from real polarization data at the level observed by the control system, both in the NMR input data and in the polarization response. In this formulation, the control variable is the microwave frequency, implemented experimentally through the EIO voltage-divider and stepper-motor settings, and the measured response is the polarization returned by the NMR analysis software, sampled with the same cadence and uncertainty as a real NMR sweep. To reproduce this behavior, the polarization dynamics are described by a generalizable phenomenological model whose parameters are inferred from experimental data spanning multiple target materials and operating conditions. The goal is not to construct a separate microscopic model for each material, but to employ a unified effective description that is sufficiently flexible to capture the dominant behavior across the target systems of interest. Within this framework, the same rate-equation structure can be extended to both spin-$\tfrac{1}{2}$ and spin-1 systems, with differences between materials and spin structure encoded through fitted model parameters. The following subsections summarize the main elements of the simulations.

\subsubsection{Rate-Equation Dynamical Model}

The time dependence of the polarization is modeled using a reduced set of coupled rate equations for a spin-$\tfrac{1}{2}$ nuclear system coupled to paramagnetic electron spins. Starting from the Leifson--Jeffries formalism \cite{jeff2}, we evolve the nuclear polarization $P_n(t)$ and electron polarization $P_e(t)$ with
\begin{align}
T_{1e}\,\frac{dP_n}{dt} &=
\left(-\frac{T_{1e}}{T_{1n}}-\frac{C\alpha}{2}-\frac{C\beta}{2}\right)P_n
+\left(\frac{C\alpha}{2}+\frac{C\beta}{2}\right)P_e,\\[4pt]
T_{1e}\,\frac{dP_e}{dt} &=
\left(\frac{\alpha}{2}-\frac{\beta}{2}\right)P_n
+\left(-1-\frac{\alpha}{2}-\frac{\beta}{2}\right)P_e
+P_0.
\end{align}
Here $T_{1n}$ and $T_{1e}$ are the nuclear and electron spin--lattice relaxation times, $C$ is the nucleus-to-electron spin ratio, and $P_0$ is the equilibrium electron polarization set by the lattice temperature and holding field. The parameters $\alpha$ and $\beta$ represent effective microwave-induced transition strengths. In the differential solid-effect picture, they correspond to two competing transition channels that, depending on microwave frequency, drive the system toward either the positive- or negative-polarization branch of the DNP response.

For fixed coefficients ($T_{1n}$, $T_{1e}$, $C$, $\alpha$, $\beta$, $P_0$), this system has the generic two-mode approach to steady state,
\begin{equation}
P_n(t)=P_n^{\mathrm{ss}} + B\,e^{-k_1 t}+D\,e^{-k_2 t},
\end{equation}
where $P_n^{\mathrm{ss}}$ is the steady-state polarization for the current microwave conditions and $k_{1,2}$ are the positive eigenvalues of the linear system. Physically, one mode corresponds primarily to the fast electron-subsystem response and the other to slower nuclear buildup or decay. This structure reproduces the observed ramp-up curves, in which polarization grows toward a plateau, and decay curves, in which polarization decreases after the system leaves optimal conditions. These are the essential time-domain signatures that the automation algorithm must interpret.

In implementation, the simulation advances $P_n$ and $P_e$ in small time steps by numerical integration so that it can respond continuously to changes in microwave frequency, beam conditions, and dose. The same core model can be executed in a stand-alone data-generator mode or in a hardware-coupled mode in which the simulated target replaces the real target but the downstream control interfaces, including frequency commands and polarization-readout timing, are unchanged.

\subsubsection{Frequency Dependence and Adjustable S-Curve}

The principal observable for frequency seeking is the steady-state nuclear polarization as a function of microwave frequency, $P_n^{\mathrm{ss}}(f)$, which experimentally has the familiar DNP S-shape with distinct positive and negative extrema. The simulation parameterizes this response in a tunable form:
\begin{itemize}
    \item The frequency dependence enters through $\alpha(f)$ and $\beta(f)$, which are treated as smooth spectral functions of frequency, such as offset Gaussian-like lobes. Their relative separation and individual widths control the overall width of the S-curve and the overlap between positive- and negative-driving regions.
    \item Rather than forcing the simplified rate equations to predict the exact S-curve from first principles, we anchor $P_n^{\mathrm{ss}}(f)$ to experiment by fitting measured steady-state points with an empirical S-curve model. The $\alpha(f)$ and $\beta(f)$ shapes are then chosen consistently with that fit so that the simulator reproduces both the correct steady-state polarization and the observed ramp-up and decay time scales as the frequency is moved.
\end{itemize}

\begin{figure*}[!htbp]
    \centering
    \includegraphics[width=\textwidth]{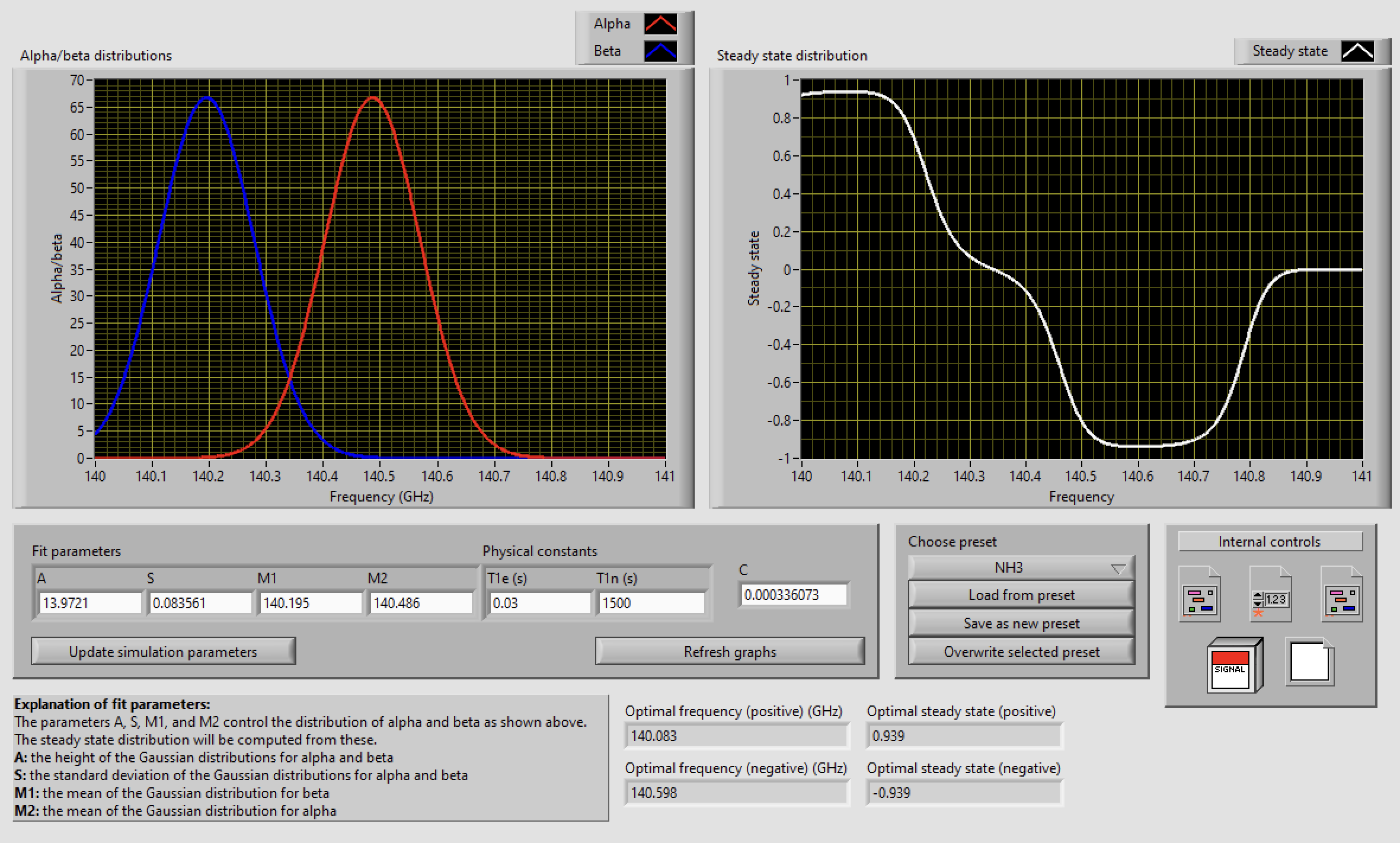}
    \caption{LabVIEW interface for the microwave-frequency-control DNP simulator. The Gaussian-like frequency distributions on the left, shown in red for negative polarization and blue for positive polarization, represent the DNP response at different frequencies and generate the corresponding S-curve on the right.}
    \label{fig:s-curve}
\end{figure*}

This approach is useful because the detailed S-curve shape depends on many experiment-specific factors, including holding field, material, radical or paramagnetic complex, temperature, microwave power, and microwave frequency. The goal of the simulation is to emulate the observed behavior well enough to stress-test and train control algorithms. Figure~\ref{fig:s-curve} shows the microwave-simulation VI\@. The Gaussian-like frequency distributions on the left, shown in red for negative polarization and blue for positive polarization, represent the DNP response at different frequencies and generate the corresponding S-curve on the right.

\subsubsection{Dose Evolution and Frequency Drift}

A defining feature of long-duration polarized-target operation is that the optimal microwave frequency is not constant. As the target accumulates dose, irradiation creates and modifies paramagnetic centers, and the locations of the positive and negative extrema drift with accumulated dose and with the history of anneals. This behavior is included explicitly by allowing the frequency-dependent response functions to evolve with dose:
\begin{itemize}
    \item The simulation tracks an accumulated dose variable $D(t)$, computed by integrating the simulated beam intensity over time using the same units and conventions as experimental dose accounting.
    \item The centers of the positive and negative branches, and, when needed, their amplitudes and widths, are permitted to depend on $D$:
    \[
    f_{+}(D),\quad f_{-}(D),\quad \text{and optionally}\quad \sigma_{\pm}(D),\ A_{\pm}(D).
    \]
\end{itemize}
As a representative parameterization for a 5~T configuration, the optimal frequencies can be written as simple exponential approaches to asymptotic values as a function of dose since the last anneal:
\begin{align}
f_{-}(D) &= 140.535 - 0.065\,e^{-3.8D},\\
f_{+}(D) &= 140.1 + 0.045\,e^{-0.38D},
\end{align}
where $D$ is expressed in the standard accumulated-dose units used during operation, such as $10^{15}\,\mathrm{e^-}/\mathrm{cm}^2$ and the corresponding units for proton beams. These relations are obtained by fitting NH$_3$ data from previous polarized-target scattering experiments. In the simulator, such relations shift the underlying S-curve and/or the $\alpha(f)$ and $\beta(f)$ spectral functions as dose accumulates, reproducing the experimentally observed frequency drift that motivates continuous retuning.

Annealing is implemented by fully or partially resetting the dose-dependent offsets, according to the chosen material parameterization, and by returning the relaxation parameters toward their pre-irradiation values. The simulator therefore reproduces the characteristic post-anneal recovery in polarization performance, followed by the gradual re-emergence of dose-induced degradation. The degradation model can also be tuned to represent different phases of target lifetime, allowing polarization decay to proceed more rapidly or more slowly as a function of accumulated dose.

\subsubsection{Beam Heating and Intensity-Dependent Depolarization}

Beam-on operation introduces additional depolarization mechanisms, primarily through beam heating and radiation damage, which reduce the achievable steady-state polarization and/or modify the relaxation characteristics. In the simulation, this behavior is implemented as an intensity-dependent modification to the relaxation and steady-state response:
\begin{itemize}
    \item The beam intensity $I(t)$, or an equivalent beam-current or flux proxy, enters as an input that modifies the effective nuclear relaxation rate and/or effective equilibrium drive term. In practice, this is implemented by replacing $T_{1n}$ with an effective value $T_{1n}^{\mathrm{eff}}(I,D)$ and, when needed, scaling the steady-state curve $P_n^{\mathrm{ss}}(f)$ by an intensity-dependent factor.
    \item The polarization does not respond instantaneously to beam changes. The rate equations ensure that any beam-induced change in the steady-state value is approached with the appropriate material-dependent time constants. Thus, turning the beam on produces a gradual decrease toward a new steady state, and turning it off produces a gradual recovery, matching the qualitative behavior observed during operation.
\end{itemize}

\begin{figure}[h!]
    \centering
    \includegraphics[width=9cm]{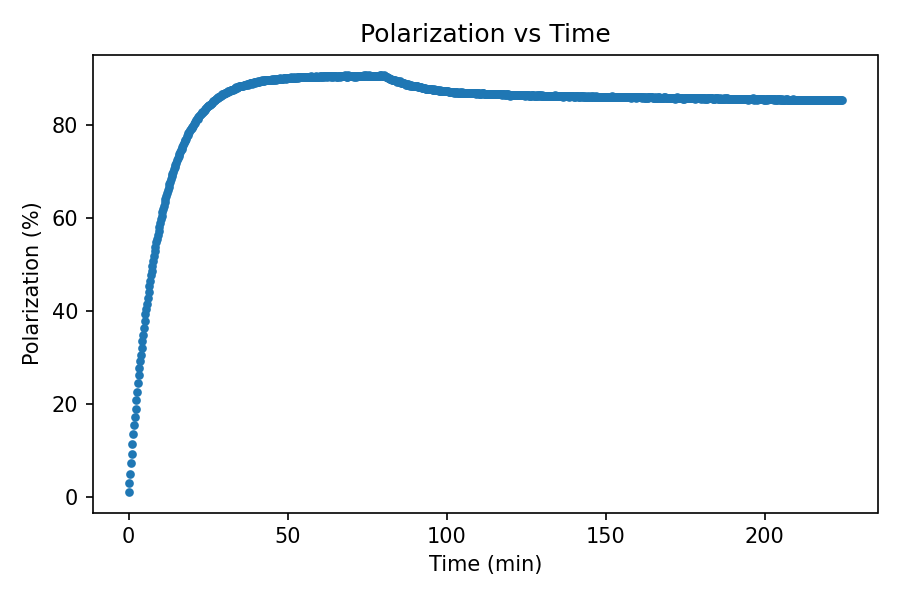}
    \caption{Simulation of NH$_3$ from the polarization-data generator, showing polarization ramp-up with a continuous electron beam turned on at about 70~min. The response exhibits a sudden drop from beam heating followed by gradual decay due to radiation damage.}
    \label{fig:sim_pol}
\end{figure}

Including beam heating is essential for automation studies because the control algorithm must distinguish between polarization losses caused by microwave-frequency detuning and those caused by changing beam conditions, including beam-intensity changes and beam trips. Figure~\ref{fig:sim_pol} shows an example of a simulated NH$_3$ polarization ramp in which a continuous electron beam is applied at approximately 70~min. The response shows an immediate polarization drop due to the additional thermal load from the beam, followed by a slower decay from radiation damage. The simulation framework includes photon, electron, and proton beam modes with configurable beam intensity and time structure, and it can also introduce random beam trips as a test condition for evaluating controller robustness.

\subsubsection{Synthetic NMR Acquisition and Monte Carlo Sampling}

The automation algorithm does not observe the true underlying polarization continuously. Instead, it receives discrete NMR-derived polarization values at the sweep cadence and with the measurement uncertainty of the NMR system. To mimic this behavior, a Monte Carlo measurement layer is applied to the continuous-time polarization dynamics:
\begin{itemize}
    \item A simulated NMR sweep schedule produces measurements at times $\{t_k\}$ consistent with the experimental sweep rate, including optional latency to represent analysis time.
    \item At each measurement time, the simulator samples the underlying polarization $P_n(t_k)$ and returns
    \[
    P_k = P_n(t_k) + \delta_k,
    \]
    where $\delta_k$ is drawn from a Gaussian distribution with zero mean and configurable standard deviation. The noise level is set to reproduce typical fluctuations of the NMR area or polarization estimator, implemented through a tunable randomness or fractional-noise parameter.
    \item The returned value is therefore a polarization value within the uncertainty of the NMR area, matching what the control code would receive during beam time.
\end{itemize}
Because the noise is generated by Monte Carlo sampling, a single control strategy can be evaluated over many pseudo-experiments to quantify robustness to measurement noise, sweep cadence, and stochastic perturbations.

\subsubsection{Material-Dependent Parameterizations and Validation}

The simulator is parameterized by target material and radical or spin concentration. For each material, including ND$_3$, NH$_3$, deuterated butanol, and LiH, we determine a consistent parameter set that includes:
\begin{itemize}
    \item a fitted steady-state S-curve $P_n^{\mathrm{ss}}(f)$ and its dose evolution;
    \item relaxation parameters $T_{1n}$ and $T_{1e}$, including dose and beam dependence when required;
    \item the effective spin-concentration parameter $C$; and
    \item the frequency-dependent transition functions $\alpha(f)$ and $\beta(f)$ that reproduce observed ramp-up and decay behavior.
\end{itemize}
These parameters are extracted by fitting dedicated experimental studies of polarization ramp-up and decay curves, both beam-off and beam-on, as well as longer-term degradation trends with accumulated dose. The resulting tuned simulations reproduce DNP performance under beam-time conditions within the experimental uncertainties of the measurements used for tuning. The synthetic datasets therefore include the coupled effects most relevant for automation: frequency drift, finite ramp-up and decay time constants, beam-induced depolarization, measurement cadence and latency, and realistic NMR noise that can be adjusted for testing.

\subsection{Automated Control}

The primary purpose of the automated control system is to ramp up the target polarization and maintain it during beam time without continuous human oversight, while also providing alarms for operating conditions that do not produce the expected maximum polarization. During ramp-up, the algorithm must identify the frequency that maximizes the polarization buildup rate. After the polarization approaches its maximum, the maintenance algorithm must continue to adjust the frequency as the beam changes the composition of the target paramagnetic complex. Although polarization inevitably decreases with accumulated dose, the goal of the maintenance algorithm is to preserve the highest achievable polarization through small compensating frequency adjustments.

\subsubsection{Basic Automation}

The automated frequency-control function is integrated into the microwave VI\@. The VI receives real-time polarization values from the NMR computer, which is connected to the target-system network. It records five consecutive polarization measurements and calculates their average value. It also determines the rate of change of polarization at each data point and computes the corresponding average rate over the same five-point interval, as given by Eq.~\eqref{equ:pol_rate}. This procedure is repeated continuously for each new set of five consecutive polarization measurements.

The control algorithm operates in two modes. The first mode is based on the polarization rate. If the average polarization rate is lower than the previous average rate, polarization growth has slowed, and the system rotates the motor by two steps in the direction opposite to the previous movement. Otherwise, the motor advances two steps in the same direction as the previous movement. Two motor steps correspond to approximately a 2~MHz change in the EIO output frequency.

\begin{equation}
P_{\mathrm{rate}}=\frac{\frac{P_4-P_3}{t_4-t_3} + \frac{P_3-P_2}{t_3-t_2}+ \frac{P_2-P_1}{t_2-t_1}+\frac{P_1-P_0}{t_1-t_0}}{4}
\label{equ:pol_rate}
\end{equation}
where $P_n$ denotes the measured target polarization at timestamp $t_n$.

In the second mode, the same decision procedure is applied to the average polarization value calculated from the five data points rather than to the polarization rate. During ramp-up, the automatic control function operates in polarization-rate mode because the rate of polarization change is significant. After maximum polarization is reached, the control algorithm switches to average-polarization mode. In this regime, where polarization variations are comparatively small, the controller relies on the averaged polarization value to maintain stable operation near the highest attainable polarization. Both modes operate for either negative or positive polarization, as selected in the VI.

\begin{figure}[h!]
    \centering
    \includegraphics[width=10cm]{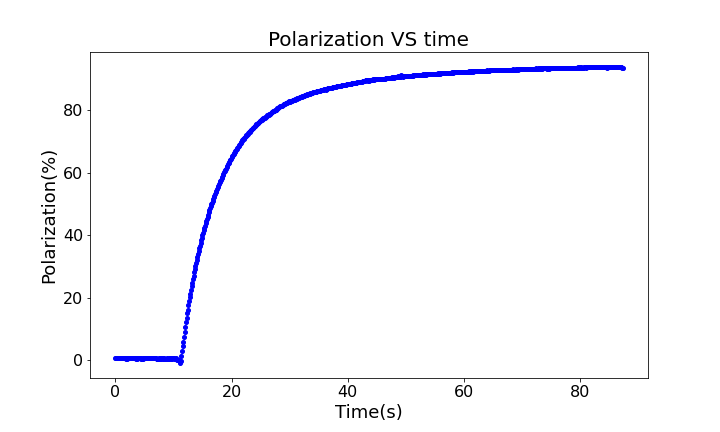}
    \caption{Positive polarization achieved in NH$_3$ during the final commissioning run using the basic heuristic automation.}
    \label{fig:positive_pol}
\end{figure}

At the beginning of the polarization process, the microwave-frequency-seeking algorithm is initialized near a known operating frequency to reduce seek time and ensure reliable convergence. The starting frequencies are set in the VI and can be adjusted as needed. The system also records previously identified optimal frequencies, so it does not need to restart the search from scratch unless the material has been annealed or replaced. To test the frequency-seeking speed of the algorithm for fresh NH$_3$, the initial frequency for positive polarization was set to 140.000~GHz, allowing the algorithm to perform a meaningful search. During automated frequency seeking, the optimum frequency for positive polarization was found in roughly 20~s to be approximately 140.14~GHz, while the optimum frequency for negative polarization was found on a similar time scale to be approximately 140.43~GHz. To achieve this the NMR was set to 16 sweeps so a polarization sample could be measure nearly every second. Figure~\ref{fig:positive_pol} shows the positive polarization obtained in NH$_3$ during the commissioning run with the automation system in operation.

Overall, the automated controller performed significantly better than manual control because it continuously evaluates the polarization slope over successive sets of measurements. Performance was characterized using the standard NH$_3$ simulation. In maintenance mode, the average error rate was approximately one incorrect step for every nine correct steps, evaluated over roughly 20 repeated trials. Ramp-up performance was quantified by the number of steps required to reach within 0.05~GHz of the optimum frequency. Starting from 140~GHz, the algorithm required an average of four steps over 20 positive-polarization and 20 negative-polarization tests. In both cases, performance depends on the NMR polarization uncertainty, specifically a change in area of approximately 3\%.

\subsubsection{Reinforcement Learning}

A reinforcement-learning (RL) architecture was studied as an alternative to the basic heuristic controller for automated microwave-frequency control. A proximal-policy-optimization (PPO) agent was trained to maximize polarization by issuing continuous frequency adjustments, with the goal of improving both ramp-up convergence speed and long-term stability relative to the rule-based system, which averages five polarization measurements and adjusts the frequency in fixed $\pm 2\,\mathrm{MHz}$ steps.

Frequency tuning was formulated as a continuous-control Markov decision process. The state vector $s_t$ includes
\begin{itemize}
    \item the current microwave frequency, normalized around 140~GHz;
    \item the previous 5--10 raw polarization measurements and their first differences, or rates;
    \item the polarity flag for positive or negative polarization; and
    \item indicators of recent polarization magnitude and short-term variance.
\end{itemize}
The agent outputs a continuous action $\Delta f \in [-400, +400]~\mathrm{MHz}$, which is quantized to motor steps before application. The reward function is shaped as
\[
r_t = P_t + \beta \cdot \Delta P_t - \lambda |\Delta f_t|,
\]
where $P_t$ is the absolute polarization, $\Delta P_t$ emphasizes growth during ramp-up, and the penalty term discourages excessive adjustments. The discount factor $\gamma = 0.99$ encourages long-horizon stability.

Control is performed by a PPO agent with a shared multilayer-perceptron (MLP) backbone, using hidden layers with 256 and 128 units and ReLU activations, feeding
\begin{itemize}
    \item an actor head that produces the mean and standard deviation of a Gaussian policy; and
    \item a critic head that estimates the state value $V(s)$.
\end{itemize}
The agent is pretrained on a calibrated simulation of NH$_3$ polarization dynamics, reproducing optima near 140.14~GHz for positive polarization and 140.43~GHz for negative polarization, and fine-tuned online during experimental runs with safety constraints, including action clipping, entropy regularization, and rule-based fallback on high-uncertainty actions.

Training and validation were carried out using the Monte Carlo DNP simulations described above. In controlled simulation studies, where polarization uncertainty, thermal fluctuations, and NMR response time can be varied systematically, the RL controller provided a modest improvement over the basic automation algorithm for identifying the optimum ramp-up frequency and tracking gradual frequency drift caused by radiation damage. This behavior is expected because the agent learns the Monte Carlo environment effectively. Although the simulation reproduces NH$_3$ behavior well, spontaneous experimental effects, such as frequency-dependent power pockets in the EIO, changes in NMR tuning, or occasional outliers in the incoming data, can degrade RL performance when they are not represented in the training environment. Under nominal conditions, the heuristic and RL controllers performed similarly. Under irregular conditions, however, the simpler heuristic algorithm was generally more robust against spontaneous variations. The RL controller can be made more adaptive by training on simulations that explicitly include these effects. Initial studies indicate that significant performance gains are possible when the training environment reproduces the characteristic frequency-dependent power response of a particular EIO\@.

Controlled simulation benchmarks indicate that the learned policy can reduce the number of ramp-up steps required to reach within $\pm 0.05\,\mathrm{GHz}$ of the optimum to approximately 2--3, compared with approximately 4 for the heuristic controller. It can also decrease the maintenance-mode error rate to below 1.5 incorrect steps per 20 otherwise optimal control steps. By capturing longer temporal dependencies and adapting to noise and material variations without hard-coded averaging windows or mode switches, the RL controller offers a promising route toward improved performance in well-characterized operating conditions.

Both the heuristic controller and the simple RL algorithm can be confused by beam trips in continuous-beam mode or by microwave power pockets unless they have been specifically tuned to a particular EIO\@. Achieving adaptive robustness against these spontaneous, complex behaviors remains challenging.

\subsubsection{Unsupervised Reinforcement Learning}

Unsupervised RL was also explored as a possible route to improved automated DNP microwave-frequency control. In this approach, the agent discovers diverse and effective frequency-adjustment strategies through self-supervised interaction with the polarization dynamics, without requiring a manually engineered reward function or explicit mode-switching rules. In a reward-free pretraining phase, the agent uses intrinsic objectives, such as maximizing state-space coverage, behavioral diversity, or skill separation, to explore the frequency--polarization landscape extensively using historical or simulated NH$_3$ data. This procedure produces a repertoire of reusable behaviors that implicitly captures rapid ramp-up, peak seeking, and stable maintenance.

When deployed, these pretrained policies can adapt rapidly to real-time NMR polarization feedback, potentially with minimal online fine-tuning or with a simple downstream objective that directly maximizes polarization. This method has not yet been used in a real experimental setting, but it may reduce the present rate of erroneous control steps, accelerate convergence to the optimum frequencies, such as approximately 140.14~GHz for positive polarization and 140.43~GHz for negative polarization, and improve robustness to noise, power pockets, beam trips, and material variations relative to the alternatives. Further studies are in progress to incorporate additional control channels, including the variable microwave attenuator, stepper motor, and power supply, together with temperature sensors near the target, the main-flow response produced by the microwave heat load, and polarization information from the NMR system. More detailed simulations with additional response features are also being developed.

\subsubsection{Integration of EIO Power-Supply Control}

A natural extension of the automated microwave control system is to include the EIO power supply in the control loop in addition to mechanical cavity tuning. Although the cathode voltage determines the broader operating region of the EIO, our initial study focused on anode-voltage control as a fine actuator around a fixed operating point. In an initial control test, the anode voltage was set to 4.44~kV with the EIO operating at 140.18~GHz. Around this point, deliberate anode-voltage changes of 1--4\% produced measurable microwave-power changes of approximately 0.1--0.5~W, inferred from changes in the main-flow helium boil-off. The same anode adjustments also shifted the EIO frequency by several 10~MHz. These measurements show that the anode supply provides a second continuously variable control channel coupled to both output frequency and delivered RF power. Software control of a modern CPI power supply can be integrated directly into a LabVIEW- or Python-based automation system; however, the interface and control logic can vary significantly among power-supply models.

When combined with stepper-motor adjustment of the EIO cavity size, anode-voltage control provides finer optimization than cavity tuning by rotation alone. The cavity position moves the system across the broader EIO tuning curve, while the anode voltage enables smaller corrections around a chosen operating point. Using the NMR polarization response together with the main-flow signal as feedback, the controller can therefore regulate both microwave frequency and RF power rather than frequency alone. This capability is important because the EIO response is not uniform across frequency: local regions of reduced output power, or power pockets, can cause a nominally favorable frequency setting to produce suboptimal polarization. Joint control of cavity position and anode voltage allows the system to avoid such regions and adapt to the characteristic response of a particular EIO\@.

From the AI-control perspective, this becomes a multivariable optimization problem. The state is extended to include the measured microwave frequency, stepper-motor position, anode voltage, variable microwave attenuator position, recent NMR polarization history, and main-flow response, while the action space becomes $(\Delta x_{\mathrm{motor}}, \Delta V_{\mathrm{anode}})$. A reinforcement-learning or model-based controller can then optimize a reward that favors large polarization and rapid ramp-up while penalizing excess cryogenic load, unstable operating points, or unnecessary actuator motion. In this framework, the controller can learn to avoid frequency regions with poor power transmission, provide finer frequency placement than is possible with cavity rotation alone, and deliberately dither the microwave frequency over a narrow bandwidth to better polarize materials with broad Larmor-frequency distributions. These initial tests indicate that integrating the EIO power supply into the AI-control system is a promising next step toward joint optimization of microwave frequency, RF power, and target polarization.

\section{Conclusion}
\label{sec:conclusion}

This work presented the SpinQuest microwave system and its automation framework for dynamic nuclear polarization of irradiated solid targets under high-field, cryogenic, and high-radiation conditions. A $\sim$140~GHz EIO, coupled to the target through a low-loss waveguide chain, was equipped with motorized cavity tuning, remote diagnostics, and LabVIEW-based supervisory control. The measured relation between tuning-shaft position and output frequency was highly reproducible over repeated scans, enabling a reliable position--frequency lookup table and repeatable frequency placement. Together with continuous readback of polarization, beam dose, and microwave-system temperatures, this system provided a practical closed-loop platform for remote operation in the SpinQuest environment.

To support controller development in a controlled and repeatable manner, a Monte Carlo DNP digital twin of the target system was constructed. The simulation reproduces the key behaviors that determine frequency-control performance: the steady-state DNP S-curve, finite buildup and decay dynamics following changes in microwave conditions, beam-heating-induced polarization loss, anneal recovery, and dose-dependent drift of the optimum microwave frequency associated with irradiation-induced evolution of the paramagnetic complex. By presenting the same frequency-control input and NMR-like polarization output seen by the real automation system, this framework provides a useful environment for training, benchmarking, and stress-testing candidate control strategies across different target materials and operating conditions.

Within this framework, a heuristic controller based on recent polarization history was implemented directly in the microwave-control VI and deployed for NH$_3$ operation. In commissioning and simulation studies, the automated system successfully identified and tracked the positive- and negative-polarization optima near 140.14 and 140.43~GHz, respectively, while reducing the need for manual intervention during ramp-up and maintenance. In the standard NH$_3$ simulation, the controller reached within 0.05~GHz of the optimum in approximately four steps on average when starting from 140~GHz, and it maintained the correct tuning direction with substantially better performance than manual retuning. Reinforcement-learning controllers were also investigated. Although they showed modest gains under nominal simulated conditions, their practical performance depended strongly on the fidelity of the training environment; under unmodeled experimental irregularities, the simpler heuristic controller was generally more robust.

An important outcome of this study is that useful automation does not require an overly complicated control architecture, provided that the diagnostics are reliable and the system model captures the dominant physics seen by the controller. At the same time, the present work identifies a clear path toward more capable multivariable optimization. Initial tests with anode-voltage control show that the EIO power supply provides an additional actuator coupled to both microwave frequency and delivered RF power. Future control systems can therefore combine cavity tuning, power-supply control, variable attenuation, cryogenic observables, and beam information to jointly optimize frequency and microwave power. Such an approach should improve sustained target polarization, reduce cryogenic load associated with unnecessary microwave power, and enhance the figure of merit of future polarized-target experiments.

\balance
\bibliographystyle{IEEEtran}
\bibliography{spinquest_cryo}

\end{document}